\documentclass[twocolumn,prb,multicol,amsmath,amssymb]{revtex4}
\usepackage[dvips]{graphicx}
\usepackage{graphicx}
\usepackage{dcolumn}
\usepackage{bm}
\usepackage{graphics}
\usepackage{epsfig,color}
\usepackage[normalem]{ulem} 
\usepackage{soul}

\newcommand{\be}{\begin{equation}}
\newcommand{\ee}{\end{equation}}
\newcommand{\bea}{\begin{eqnarray}}
\newcommand{\eea}{\end{eqnarray}}

\begin{document}
\title {Spin squeezing: Thermal behaviour and distribution on excited states}
\author{Saeed Mahdavifar$^{1}$}
\author{Hadi Cheraghi$^{2,3}$}
\author{Kourosh Afrousheh$^{4}$}
\affiliation{$^{1}$Department of Physics, University of Guilan, 41335-1914, Rasht, Iran}
\affiliation{$^{2}$Computational Physics Laboratory, Physics Unit, Faculty of Engineering and Natural Sciences, Tampere
University, Tampere FI-33014, Finland}
\affiliation{$^{3}$Helsinki Institute of Physics, University of Helsinki FI-00014, Finland}
\affiliation{ $^{4}$Department of Physics, Kuwait University, P. O. Box 5969, 13060-Sadaf, Kuwait}
\begin{abstract}

We investigate the spin-squeezing behavior under thermal effects in a one-dimensional transverse field XY model with spin-1/2. The exact solution of the model helps us to compute the spin-squeezing parameter as a function of temperature and also in all excited states with higher energy than the ground state. We find that below the thermal factorized field, $h_f(T_{co})$, there is no transition temperature. At the thermal factorized field, a transition from a thermal squeezed state to an unsqueezed state occurs at a specific temperature called the “coherent temperature”. Interestingly, we show that the finite temperature can create squeezed states from a state which at zero temperature is a coherent state. To complete our study, we also analyze the variation of the spin-squeezing parameter in the excited states and provide a behavioral analysis of the thermal spin-squeezing parameter.
\end{abstract}
\maketitle

\section{Introduction}\label{sec1} 

Spin squeezing is a quantum effect that makes one direction of the total spin of a system of particles, such as atoms or electrons, more precise and less noisy [{\color{blue}\onlinecite{1971Radcliffe,1993Kitagawa,Int9,Int10}}]. This means that the system can effectively detect small rotations in that direction. Spin squeezing has many uses in quantum physics, such as quantum entanglement [{\color{blue}\onlinecite{I10,I11,I12,I13,I14,I15,I16}}],  which is a quantum phenomenon where the quantum state of each particle cannot be described independently of the others, even when the particles are separated by large distances, and quantum phase transition [{\color{blue}\onlinecite{12n, 13n, 14n, 15n, 16n, 17n,18n,18n0, 18n1, 19n, 20n, 20nn}}],  which is a type of phase transition that occurs at absolute zero temperature and is driven by quantum fluctuations rather than thermal energy.
 In optical systems, precise knowledge of the phase of light is crucial for various applications. However, direct measurement of the light's phase is not possible; instead, phase estimation techniques are employed  [{\color{blue}\onlinecite{Th-5}}]. The accuracy of usual phase estimation is limited by the standard quantum limit. Hence, spin squeezing also is applied in quantum metrology [{\color{blue}\onlinecite{I5,I6,I7,I8,I9}}], which is the field of study that uses quantum theory to make high-resolution and highly sensitive measurements of physical parameters.
In addition, spin squeezing has been the subject of many interesting experimental researches in recent years [{\color{blue}\onlinecite{Thex-0, Thex-1,Thex-2,Thex-3, Thex-4, Thex-5, Thex-6}}]. 

Thermal spin-squeezing is a phenomenon where the quantum fluctuations of one component of a collective spin operator are reduced below the standard quantum limit, which is the minimum uncertainty achievable with classical states, in a group of quantum spins that are in thermal equilibrium with their environment. Thermal spin-squeezing has attracted a lot of attention in recent years, and several studies have explored its properties and applications [{\color{blue}\onlinecite{Th-1, Th-2, Th-3, Th-4, Th-5, Klaers}}]. For example, some studies have investigated how the spin squeezing parameter (SSP) is influenced by the phonons, which are the quanta of the bosonic modes, and how it depends on the temperature, the system size, and the integrability breaking perturbations [{\color{blue}\onlinecite{Th-1}}]. Some have proposed a method to prepare squeezed thermal states at a controlled temperature [{\color{blue}\onlinecite{Th-2}}]. In addition, it is indicated how the range of fields and temperatures for a thermal-equilibrium state can improve the resolution in an NMR experiment and probe relevant parameters of the quadrupole Hamiltonian [{\color{blue}\onlinecite{Th-3}}]. By applying a magnetic field gradient and a microwave pulse, it is shown that the thermal gas of rubidium atoms,  which are spin-1 particles, can be prepared in a coherent superposition of spin states, and then evolve under a one-axis twisting Hamiltonian, which generates spin squeezing [{\color{blue}\onlinecite{Th-4}}]. Further, using a vibrating nano-beam driven by squeezed electronic noise, it is illustrated that in coupling the working medium of a nanoscale heat engine to a squeezed reservoir, one can generate work beyond the Carnot’s limit [{\color{blue}\onlinecite{Klaers}}].

\begin{figure}
\centerline{\includegraphics[width=1.1\linewidth, height=0.22\textheight]{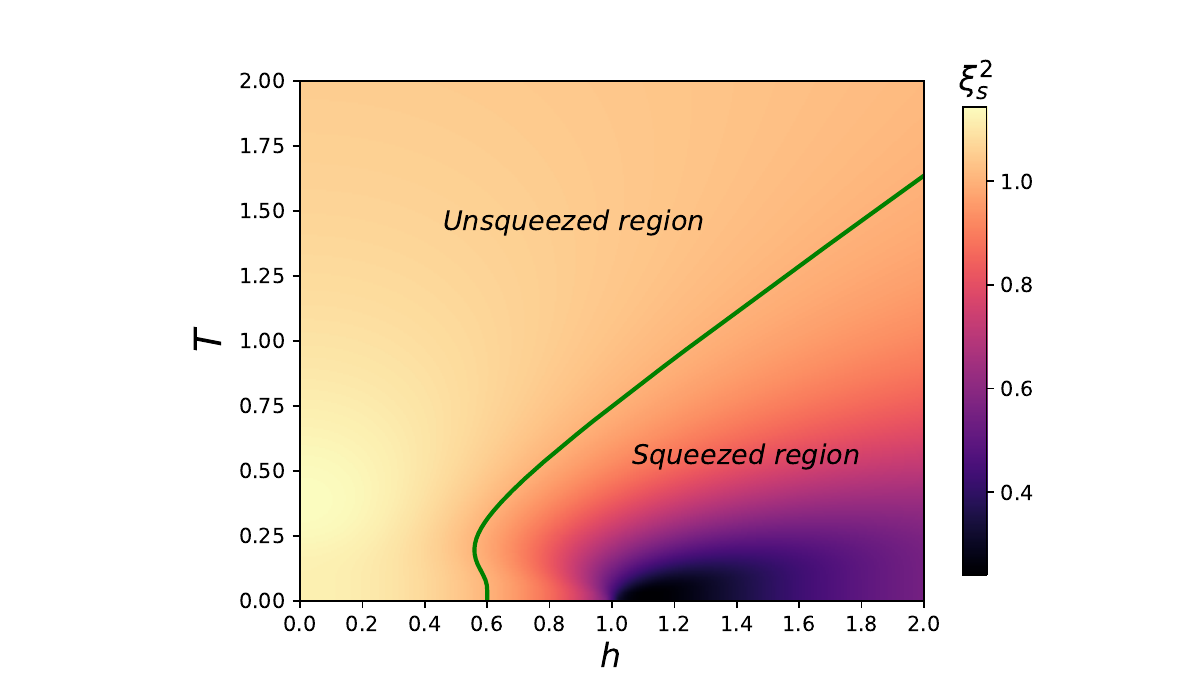}}
\caption{(color online). The density plot of the thermal SSP for a chain size $N=200$ with $\delta=0.8$. The green line is where the transition from unsqueezed to squeezed thermal states happens.}
\label{Fig1}
\end{figure}

One-dimensional (1D) quantum spin systems exhibit numerous non-classical properties related to spin squeezing. Specially, the 1D spin-1/2 transverse field (TF) XY chain model and related systems are a widely studied topic in this way [{\color{blue}\onlinecite{Th-7,Th-7-1,Th-80,Th-8,Th-9,Th-6,Th-10}}]. Its ground state exhibits two phases at zero temperature: a ferromagnetic (FM) phase with a broken $Z_2$ symmetry in the infinite system size limit, and a paramagnetic (PM) phase. Here, we examine the thermal effects on the SSP in a 1D TF XY model with spin-1/2. First, we use the fermionization technique to diagonalize the Hamiltonian of the system and obtain its eigenvalues and eigenvectors.  Then the thermal SSP is calculated exactly. The results of the thermal behaviours suggest that there is a thermal transition from unsqueezed to squeezed regions at a given temperature. Consequently, the factorized field will be dependent on this temperature. By choosing different initial states, we will investigate the resilience of the squeezing parameter in the presence of temperature.
 We also compute the densities of both squeezed and unsqueezed states, along with their distribution across the spectrum of excited states. The investigation into the distribution of spin squeezing delves into its variation within a quantum system when the system occupies an excited state—that is, a state possessing energy levels above the ground state. Our findings indicate that the densities of squeezed and unsqueezed states are highest at the middle of the excited spectrum, with the majority of excited states aggregating within the nonzero domain of the SSP.

The paper is organized as follows. In the next section, we introduce the model and employ the fermionization approach to derive the system's spectrum. The calculation of the thermal spin squeezing is presented in section III. In Section IV, we present our findings regarding the thermal behavior and density of the SSP in all excited states. Finally, in Section V, we provide our conclusions and a summary of the results.


\section{The model}

As a benchmark model, we consider the Hamiltonian of the 1D spin-1/2 XY model in the presence of a TF defined by
\begin{eqnarray}\label{eq1}
{\cal H} =&-&  \sum\limits_{n = 1}^N {\left[ {(1\! + \!\delta )S _n^x S_{n \!+ \!1}^x + (1 \!- \!\delta ) S_n^y S_{n \!+\! 1}^y} \right]} \nonumber \\
&- &h\sum\limits_{n \!= \!1}^N {S_n^z},
\end{eqnarray}
with $N$ the number of sites on the chain, $S_n^{\vartheta}$ as spin operators on-site $n$ represented by Pauli matrices $S_n^{\vartheta}=\sigma_n^{\vartheta}/2$ for $\vartheta = x, y, z$,  $\delta$ and $h$ denote the anisotropy parameter and magnitude of the TF, respectively. We here consider the case of periodic boundary conditions, $S_{n+N}^\vartheta=S_n^\vartheta $. In addition, we restrict the anisotropy parameter to the interval $(0,1]$.  The model in question stands out as one of the select quantum many-body systems that can be solved exactly, facilitating an in-depth analysis of quantum phase transitions and associated critical phenomena. It acts as a foundational framework for investigating entanglement characteristics and quantum information theory elements, including quantum correlations [{\color{blue}\onlinecite{Osterloh2002, Vidal, Maziero, E9-00}}]. Moreover, it effectively emulates the magnetic behaviors of specific real-world materials, thus aiding in the comprehension of their magnetic properties. Notably, magnetic compounds like $\rm Cs_2CoCl_4$ are accurately characterized by the XY model [{\color{blue}\onlinecite{E9-0,E9-1}}].

In the thermodynamic limit $N \rightarrow \infty$, and at zero temperature, the model exhibits a quantum phase transition at $h_c \!= \!1$, from an FM phase ($h \!< \!1$) to a PM phase ($h\!>\!1$) [{\color{blue}\onlinecite{E9,E10,E11}}]. 
On the circle, $h_f^2+\delta ^2=1$, the wave function of the ground state is collapsed into a product of single spin states.   This line is referred to as the factorized line, at which the ground state becomes an eigenstate of one of the total spin operators, thereby facilitating a coherent state [{\color{blue}\onlinecite{Radcliffe, Kurmann, Adesso}}]. Therefore, the factorized field, $h_f$, is dependent on the Hamiltonian's parameters. At nonzero temperature, this field is also dependent on the temperature, $h_f(T_{co})$, which we name the “thermal factorized field”. 
We find that for TFs below the $h_f(T_{co})$, the system stays in the unsqueezed thermal state, and no transition occurs. At the thermal factorized field, the system undergoes a transition from a squeezed thermal state to an unsqueezed thermal state at a specific temperature called the “coherent temperature”, $T_{co}$.  In fact, the coherent temperature refers to a specific finite temperature at which the system's thermal state achieves coherence, characterized by the condition $\xi_s^2 = 1$ where $\xi_s^2$ is the SSP. 
Our results reveal a proximate relationship for the thermal factorized field as a function of $T_{co}$ as
\begin{eqnarray} \label{eq2}
h_f(T_{co}) \approx h_f+aT_{co}+b\tanh(cT_{co}),
\end{eqnarray}
where the constant values $a,b,c$ are dependent on the values of the Hamiltonian's parameter, i.e., $\delta$.
As we can see from Fig.~(\ref{Fig1}), dependent on the temperature, this transition can appear for $h_f(T_{co})$ below and above the critical field $h_c=1.0$. Remarkably, there is also a small region where $h_f(T_{co})<h_f$. Here, $T_{co}$ appears double valued in a narrow range versus $h$.

The Hamiltonian in (\ref{eq1}) is integrable and hence has an exact solution by fermionization technique [{\color{blue}\onlinecite{LSM}}]. 
Using the Jordan-Wigner  transformation, 
\begin{eqnarray}\label{eq3}
\sigma _n^ +  &=& \prod\limits_{l = 1}^{n - 1} {(1 - 2c_l^\dag {c_l})} {c_n}~;~\sigma _n^ -  = \prod\limits_{l = 1}^{n - 1} {(1 - 2c_l^\dag {c_l})} c_n^\dag \nonumber\\
\sigma _n^z &=& 2c_n^\dag {c_n} - 1
\end{eqnarray}
where, $\sigma_ n^\pm = (\sigma_n^x \pm i\sigma_n^y)/2$, and $c_n^\dag $ and $c_n$ are the fermionic creation and annihilation operators, followed by a Fourier transformation $c_n=(1/\sqrt{N})\sum_k \exp[ikn]c_k$, the Hamiltonian turns to the form of
\begin{eqnarray}\label{eq4}
H= \sum_{k>0} {\cal C}^\dag {\mathbb{H} }_k  {\cal C},
\end{eqnarray}
where ${\cal C}^\dag =(c_k^\dag,c_{-k})$ and ${\mathbb{H}}_k={\overrightarrow d _k}.\overrightarrow \sigma  $
with the Bloch vector ${\overrightarrow d _k}=(d_x,d_y,d_z)$  and $\overrightarrow \sigma=(\sigma^x,\sigma^y,\sigma^z)$  that are the Pauli matrices. Here $d_x=0$, $d_y=-\delta \sin (k)$, and $d_z =-(\cos(k)+ h)$.  However, performing Bogoliubov transformation  $c_k = \cos (\theta _k)\mu  _k + i\sin (\theta _k)\mu _{ - k}^\dag $ yields the quasiparticle Hamiltonian
\begin{eqnarray}\label{eq5}
H=\sum\limits_{k} \Lambda_k (\mu_k^\dag \mu_k -1/2),
\end{eqnarray}
with energy spectrum $\Lambda _k = \sqrt{d_y^2 + d_z^2}$. The summation runs over $k=2\pi m/N$, with $m=0,\pm 1,...,\pm \frac{1}{2}(N-1) \ [m= 0, \pm 1,..., \pm (\frac{1}{2}N-1), \frac{1}{2}N]$ for $N$ odd [$N$ even] [{\color{blue}\onlinecite{Dong}}].
 Here, without loss of generality, we set $\delta = 0.8$, which results in the emergence of the factorized field at $h_f = 0.6$ at zero temperature.


\section{Thermal Spin Squeezing} 

 The total spin operators are defined as $J_{\alpha} = \sum_{n=1}^{N} S_n^\alpha$ for $\alpha=x,y,z$. The uncertainty relations for these operators stem from their commutation relations $[J_{\alpha}, J_{\beta}] = i \varepsilon_{\alpha \beta \gamma} J_{\gamma},$ where $\varepsilon_{\alpha \beta \gamma}$ represents the Levi-Civita symbol. This leads to the spin-uncertainty relation $ (\Delta J_{\alpha})^2 (\Delta J_{\beta})^2 \geq |\langle J_{\gamma} \rangle|^2/4$, with $\Delta J_{\alpha} = \sqrt{\langle J_{\alpha}^2 \rangle - \langle J_{\alpha} \rangle^2}$.
The SSP is derived from the Heisenberg uncertainty principle for the total spin components, imposing a minimum on the variances' product. 
In other words, when one of the fluctuations on the left-hand side satisfies $(\Delta J_{\alpha})^2 < |\langle J_{\gamma}\rangle|/2$, implying a squeezing parameter $\xi^2 = 2(\Delta J_{\alpha})^2/|\langle J_{\gamma}\rangle | \ ( \alpha \neq \gamma$).
The Kitagawa-Ueda [{\color{blue}\onlinecite{1993Kitagawa}}] and the Wineland [{\color{blue}\onlinecite{Int9}}] parameters are two SSP metrics. The former is ideal for spin systems with a clear mean spin direction and a large number of particles, while the latter fits systems sensitive to $SU(2)$ rotations with a small number of particles. The Kitagawa-Ueda SSP is defined by
\begin{equation}{\label{eq6}}
\xi_s^2 = \frac{4(\Delta J_{\vec{n}_{\perp}})^2}{N}.
\end{equation}
In this context, $\xi_s^2$ is the SSP, $\vec{n}_{\perp}$ is the axis orthogonal to the mean spin direction $\vec{n}_0$, and the variance $(\Delta J)^2$ is minimized over all directions $\vec{n}_{\perp}$.  
It is desirable that this equation signals a squeezed state when $\xi_s^2 < 1$ and a coherent state when $\xi _s^2 =1$. 
A coherent state has a minimal uncertainty state that describes the maximum coherence and classical behavior. Therefore, a thermal coherent state generalizes the notion of a coherent state to finite temperatures.

For our model, the mean-spin direction is along the $z$ direction. Therefore, one can read $J_{\vec{n}_{\perp}} = \cos (\Omega )J_x + \sin (\Omega)J_y$, where $\Omega$ must be chosen so that minimizes the SSP. Our calculations reveal that
\begin{eqnarray}\label{eq7}
\xi_s^2= 1+ 2 \sum\limits_{n = 1}^{N - 1} ( G_n^{xx} + G_n^{yy})-2|\sum\limits_{n = 1}^{N - 1} ( G_n^{xx} - G_n^{yy})|,
\end{eqnarray}
where  $G_n^{pq}$ are  two-point correlation functions defined by $G_n^{pq } := \langle S_1^{p} S_{1+n}^{q}\rangle$ with $p,q=x,y,z$. For details see the Appendix.
The calculation of $G_n^{pq}$ involves computing strings of operators of the forms 
\begin{eqnarray}\label{eq8}
G_n^{xx} &=& \frac{1}{4}\langle {{B_1}{A_2}{B_2} \cdots {A_n}{B_n}{A_{n + 1}}}\rangle, \\
G_n^{yy} &=& \frac{{{{( - 1)}^n}}}{4}\langle {{A_1}{B_2}{A_2} \cdots {B_n}{A_n}{B_{n + 1}}}\rangle, \nonumber
\end{eqnarray}
where $A_j=c_j^\dag + c_j$, $B_j=c_j^\dag - c_j$ in the  Jordan-Wigner basis. The nonlocal nature of the Jordan-Wigner transformation makes $G ^{pq}_n$ nontrivial. To this end, one can rely on Toeplitz determinants [{\color{blue}\onlinecite{LSM,Barouch2}}] resulting in an exact expression for $G ^{pq}_n$. 
On the other hand, in order to access the thermal behavior, in the eigenbasis of $\left\{ |0_k0_{ - k}\rangle ,c_{k}^{\dag} c_{-k}^{\dag} | 0_k 0_{ - k}\rangle ,c_{k}^{\dag}| 0_k0_{ - k}\rangle ,c_{-k}^{\dag}| 0_k 0_{ - k}\rangle \right\}$, one can write the Hamiltonian as $H=\sum_{k>0} H_k$, where 
\begin{eqnarray}\label{eq9}
H_k = \left[ {\begin{array}{*{20}{c}}
{d_0}&{-id_y}&0&0\\{id_y}&{d_0 + 2d_z}&0&0\\
0&0&{d_0 + d_z}&0\\0&0&0&{d_0 + d_z}\end{array}} \right], 
\end{eqnarray}
with $d_0=h$. This representation helps us to write the density matrix at the equilibrium $\rho=e^{-\beta H}/{\rm Tr}(e^{-\beta H})$ in the form of $\rho = \mathop  \otimes \nolimits_{k > 0} {\rho _k}$. It is easy to show
\begin{eqnarray}\label{eq10}
\rho _k =\frac{1}{\Theta _k} \left[ {\begin{array}{*{20}{c}}{d_{11}}&{d_{12}}&0&0\\{d_{21}}&{d_{22}}&0&0\\
0&0&{1}&0\\0&0&0&{1}\end{array}} \right],
\end{eqnarray}
with
\begin{eqnarray}\label{eq11}
d_{11/22} &=&  \cosh (\beta \Lambda _k) \pm \cos(2\theta_k)\sinh (\beta \Lambda _k),   \nonumber\\
d_{12}  &=& -d_{21}=  - i \sin(2\theta_k) \sinh(\beta \Lambda _k),       
\end{eqnarray}
and $\Theta_k = 2[1 + \cosh (\beta \Lambda _k)]$. Here 
\begin{eqnarray}\label{eq12}
\cos(2\theta_k)=\frac{d_z}{\Lambda _k} ~;~\sin(2\theta_k)=-\frac{d_y}{\Lambda _k}.  
\end{eqnarray}
Note that the average $\langle {\cal O} \rangle={\rm Tr}(\rho {\cal O})$  is taken over the thermal ensemble with  $\beta = \frac{1}{k_B T}$ where $k_B$ is the Boltzmann constant. In our calculation, we put $k_B=1$. By applying Wick theorem to express correlators of fermion operators as Pfaffians [{\color{blue}\onlinecite{Fubini}}], one can obtain
the elements of the Pfaffians through
\begin{eqnarray}\label{eq13}
\langle A_nA_ m \rangle  &=& -\langle B_nB_m \rangle=   \delta _{n,m}, \\
\langle A_nB_m \rangle  &=&\frac{1}{N}\sum\limits_{k } \tanh (\frac{\beta \Lambda _k}{2}) \cos (2\theta _k + km), \nonumber \\
\langle B_nA_m \rangle  &=&  -\frac{1}{N}\sum\limits_{k } \tanh (\frac{\beta \Lambda _k}{2}) \cos (2\theta _k - km). \nonumber 
\end{eqnarray}


\section{Results}

\begin{figure}[t]
\centerline{\includegraphics[width=1.1\linewidth]{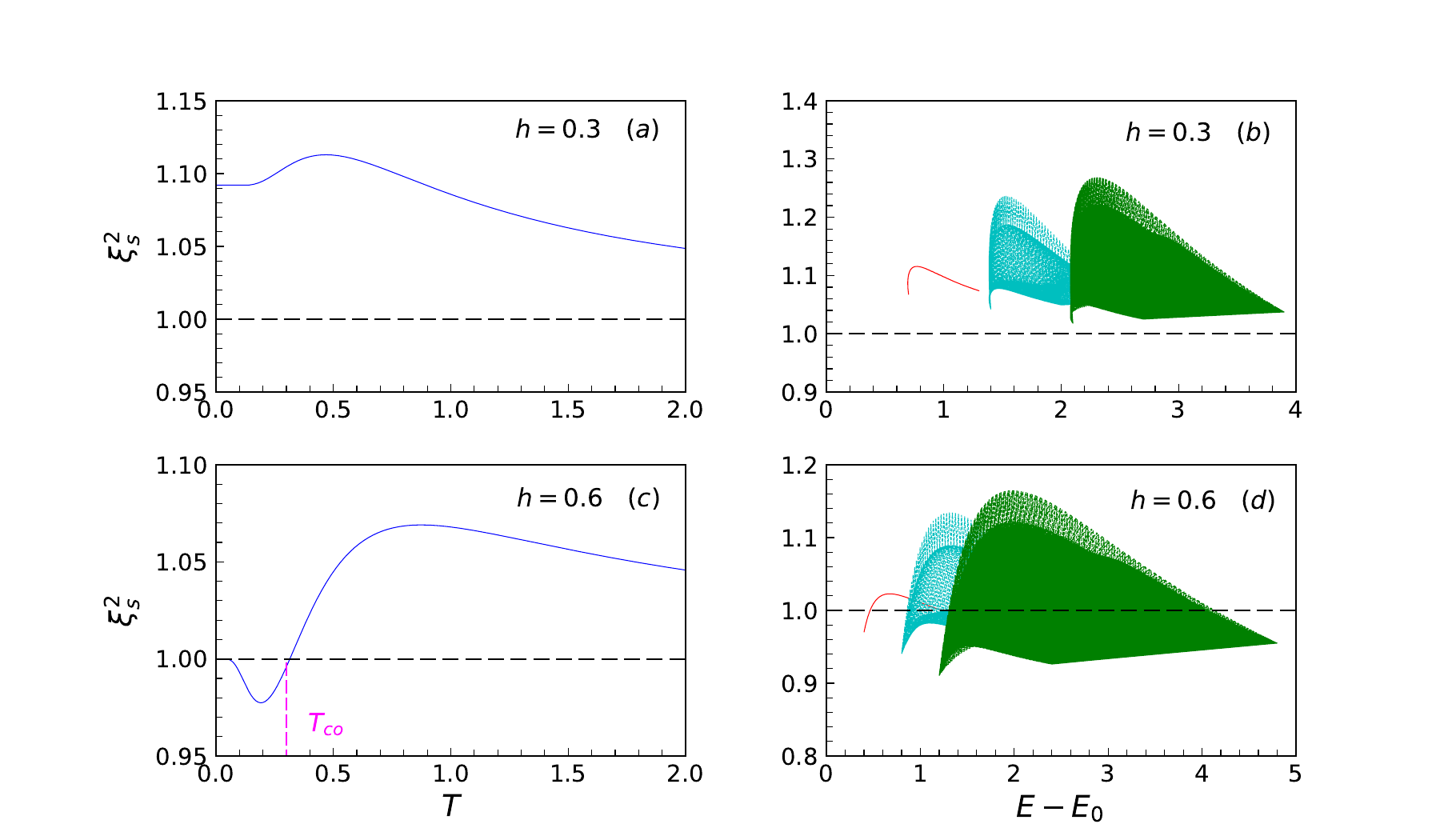}}
\caption{(color online). The temperature dependence of the SSP for different values of the TF is as (a) $h=0.3$, and (c) $h=h_f=0.6$. The vertical dashed pink lines indicate the coherent temperature, $T_{co}$. The SSP as a function of the energy difference between excited ($E$) and ground ($E_0$) states for different values of TF as (b) $h=0.3$, and (d) $h=h_f=0.6$. Here the red, cyan, and green plots correspond on the subspaces with $N_B$ as 1, 2, and 3, respectively. For all plots, the size of the chain is $N=200$ and the horizontal dashed black lines hint at the coherent line where $\xi_s^2=1.0$. }
\label{Fig2}
\end{figure}

\begin{figure}[t]
\centerline{\includegraphics[width=1.1\linewidth]{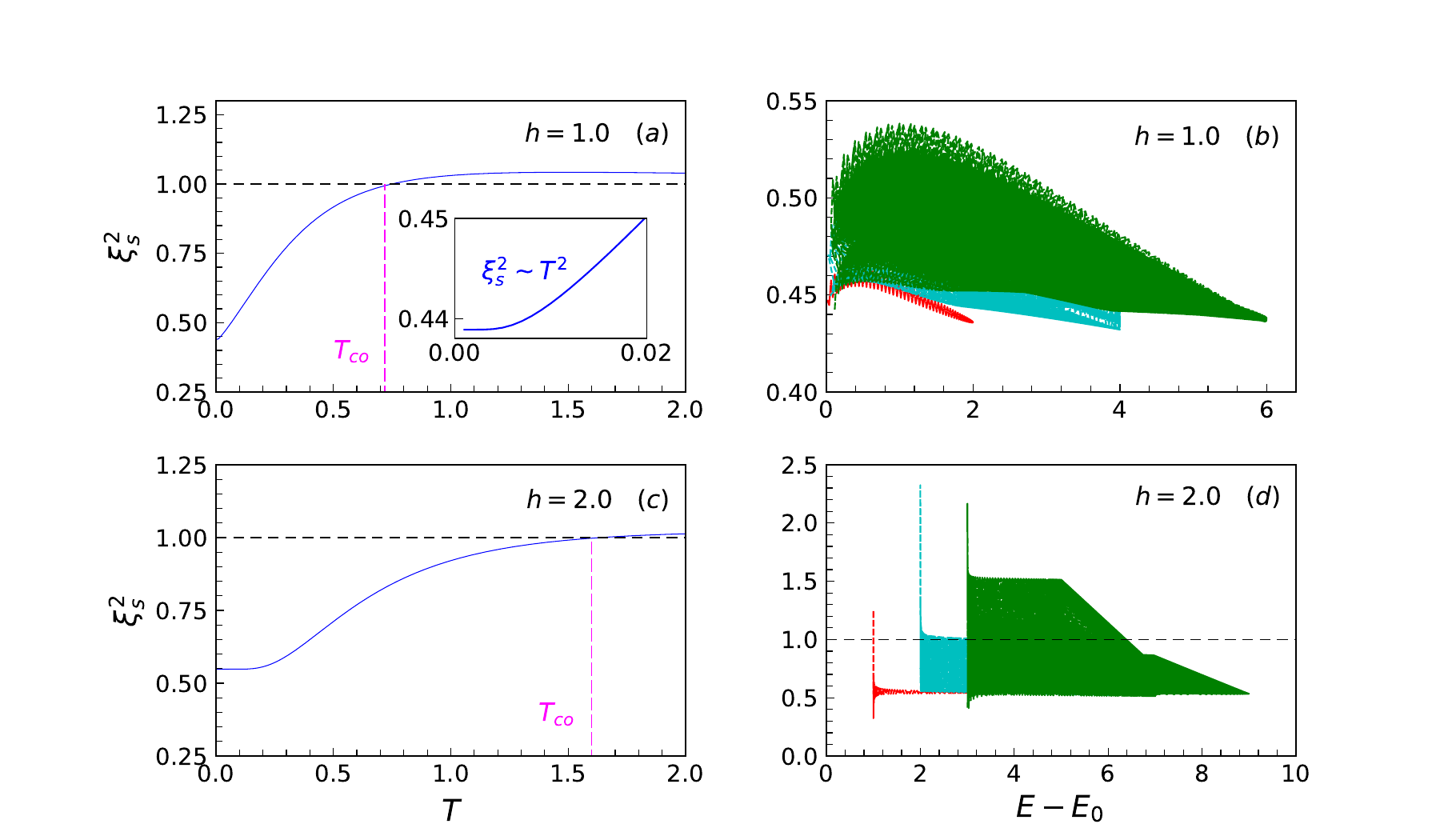}}
\caption{(color online). The temperature dependence of the SSP  for different values of the TF is as (a) $h=h_c=1.0$ and (c) $h=2.0$. In addition, the inset in (a) clearly displays the scaling of SSP for the finite temperature near zero, $\xi _s^2 \sim {T^2}$. The vertical dashed pink lines show $T_{co}$. The SSP as a function of the energy difference $E-E_0$ for different values of TF as (b) $h=h_c=1.0$, and (d) $h=2.0$. The red, cyan, and green plots belong to the subspaces with $N_B$ as 1, 2, and 3, respectively. Explicitly is viewed that at the critical point, $h_c=1.0$, the gap closes, $E-E_0=0$.  For all plots, the size of the chain is $N=200$ and the horizontal dashed black lines illustrate $\xi_s^2=1.0$. }
\label{Fig3}
\end{figure}

In this section, we report our exact analytical results which are obtained for a chain size $N=200$ and anisotropy $\delta=0.8$.  It is known that at zero temperature, the ground state is unsqueezed in the region $h<h_f=0.6$ and squeezed in the region $h>h_f$. In addition, it becomes coherent exactly at the factorized field, $h_f$. The minimum of $\xi _s^2$ occurs not exactly at the quantum critical field, $h_c=1$, but a little more than it ($h\simeq 1.1$).  In Fig.~\ref{Fig1}  we have plotted the $\xi_s^2$ as function of $h$ and $T$. As is clear, thermal states are separated into two regions, squeezed and unsqueezed states. The boundary between these two is the thermal factorized field $h_f(T_{co})$, marked by a green line, where at a very large temperature, it has a linear relationship with  $T_{co}$ ($\tanh(cT_{co}) \approx  0$ in Eq.~(\ref{eq2})).  Indeed, at finite temperatures, the system's state evolves smoothly across $\xi^2 = 1$, making $T_{co}$ defined only up to a factor of order one. Therefore, we consider $T_{co}$ as a cross-over scale rather than a sharply defined transition temperature. This perspective aligns with the inherent smoothness of the system's evolution at finite temperatures. In addition, the minimum of the SSP for the low temperature still emerges around the TF a little more than $h_c$ while for high temperatures it shifts to bigger TFs. In the following, we will study how the SSP changes with temperature for four different values of the TF as: (I) $h<h_f$, (II) $h=h_f$, (III) $h=h_c=1.0$, and (IV) $h>h_c=1$. Figures.~\ref{Fig2} and \ref{Fig3} explicitly show the results.

For $h<h_f$, Fig.~\ref{Fig2}(a), $\xi _s^2$ is constant at very low temperatures, indicating a gapped system. As the temperature increases, it first rises to a maximum, then falls again, approaching the coherent thermal state at $T\longrightarrow \infty $.  It is known that at sufficiently high temperatures, thermal fluctuations will destroy all correlations, $G^{pq}_{n}\longrightarrow 0 $, which implies that all thermal states act as coherent states with $\xi^2_s \longrightarrow  1$. 
We also look at the excited state spin squeezing and how it is distributed. This is important, as it can reveal the quantum dynamics, thermodynamics, and phase transitions of the system. It can also help us to manipulate or measure the quantum properties of excited states [{\color{blue}\onlinecite{Thex-7}}]. We study the SSP in the excited states of the system, which are labeled by the Bogoliubov fermion number $N_B$. The Hamiltonian has $\hat{N}_{B}=\sum_k  \mu_k^\dag \mu_k $ as a constant of motion, with eigenvalues $N_B={0,1,2,....,N}$. The ground state is the vacuum state with $N_B=0$ and the total energy $E_0$.  Figure.~\ref{Fig2}(b) shows the SSP versus the total energy difference between excited, $E$, and ground states in subspaces with $N_B= 1, 2, 3$. We find consistent behavior for up to $N_B=5$. The energy gap is visible in the figure. All eigenstates are unsqueezed for $h=0.3<h_f(T_{co})$, so there is no transition to the squeezed state by increasing temperature for $h<h_f(T_{co})$.  A sudden onset of increased density of data points is clearly seen.   However, the middle eigenstates of each subspace have higher $\xi _s^2$ than the edge eigenstates, which leads to a peak in the SSP's thermal behavior. This peak is reminiscent of the specific heat capacity anomaly observed in solid-state physics, known as the Schottky anomaly, which manifests as a peak in the specific heat capacity at low temperatures [{\color{blue}\onlinecite{Thex-7-0}}].

In Fig.~\ref{Fig2}(c) we study the SSP as a function of temperature for the system initialized at the factorized field $h=h_f$. As soon as the temperature rises, $\xi _s^2$ decreases, indicating thermal squeezing. Then, it reaches a minimum, followed by a maximum, and finally approaches the coherent thermal state at $T\longrightarrow \infty $. 
As an important result, as viewed, the finite temperature is able to create squeezed thermal states when the ground state at the zero temperature is kept in a coherent state.
Figure.~\ref{Fig2}(d) expresses the results $E-E_0$ for subspaces with $N_B= 1, 2, 3$.  The energy gap is evident and smaller than $h<h_f$. The low excited states in each subspace are squeezed, which causes $\xi _s^2$ to decrease with temperature. The middle states are unsqueezed and appear condensed, which causes $\xi _s^2$ to increase with temperature. There is a transition from unsqueezing to squeezing at a specific temperature, which we called it as the coherent temperature, $ T_{co}$.

In the next step, the thermal behavior of the SSP is considered at the quantum critical field $h=h_c=1$. Results are presented in  Fig.~\ref{Fig3}(a). As is clearly seen, as soon as the temperature increases, $\xi _s^2$ increases and reaches a transition to an unsqueezed thermal state at $T=T_{co}$.  The rapid increase in the thermal average exceeding one is due to the gapless spectrum and high degeneracy.  Moreover, thermal scaling and critical exponent characterize how physical quantities change near continuous phase transitions, where the system experiences a sudden change of state. We have studied the critical exponent near the quantum critical point $h=h_c$. The results are shown in the inset of Fig.~\ref{Fig3}(a). It can be seen that it scales quadratically with the temperature near the quantum critical point. One possible reason for this scaling is based on the quantum critical metrology approach [{\color{blue}\onlinecite{E19-5}}]. This approach connects the SSP to the quantum Fisher information, ${\cal F}_Q$,  as $\xi_s^2 \leq (N/{\cal F}_Q)$ [{\color{blue}\onlinecite{Smerzi2009}}],  which is a measure of the sensitivity of the quantum state to a variation in the parameter of interest. The quantum Fisher information can be written in terms of the energy gap and the specific heat of the system, which are both known to scale quadratically with the temperature near the critical point. Therefore, the SSP also scales quadratically with the temperature near the critical point. This scaling implies that the SSP is more susceptible to the temperature near the critical point than away from it. Results drown in Fig.~\ref{Fig3}(b) belongs to $\xi _s^2$ as function of $E-E_0$.  The spectrum is gapless, and the ground state has the lowest $\xi _s^2$ among all states. In addition, a sudden onset of increased density of data points is also clearly seen. To this reason, as quickly as the temperature enhances, $\xi _s^2$ grows rapidly. In addition, as is seen, by increasing the number of fermions or in subspaces with larger values of $N_B$ ($N_B \gg N$), the amount of $\xi _s^2$ increases which unveils that for large enough values of $N_B$, there are unsqueezed excited states which play the main role in unsqueezing behavior of the SSP and mentioned transition at $T_{co}$.       

Finally, we study the SSP as a function of temperature for the system in the region $h>h_c$. We choose $h=2.0$ as an example and the results are depicted in Fig.~\ref{Fig3}(c). $\xi _s^2$ has a plateau at low temperatures, indicating a gapped system. It then increases smoothly and reaches a transition to an unsqueezed thermal state at a certain temperature. In Fig.~\ref{Fig3}(d), $\xi _s^2$ versus the excited state energy is plotted. The energy gap is clearly visible.  A narrow peak at the lowest energies is observed in every subspace. Additionally, a plateau-like region with the same maximum value spans a wide range, encompassing both squeezed and unsqueezed states.  The high excited states have similar SSP values to the ground state. However, there are enough unsqueezed states in the middle of the spectrum that cause the transition to unsqueezing by increasing the temperature.

\begin{figure}[t]
\centerline{\includegraphics[width=1.15\linewidth]{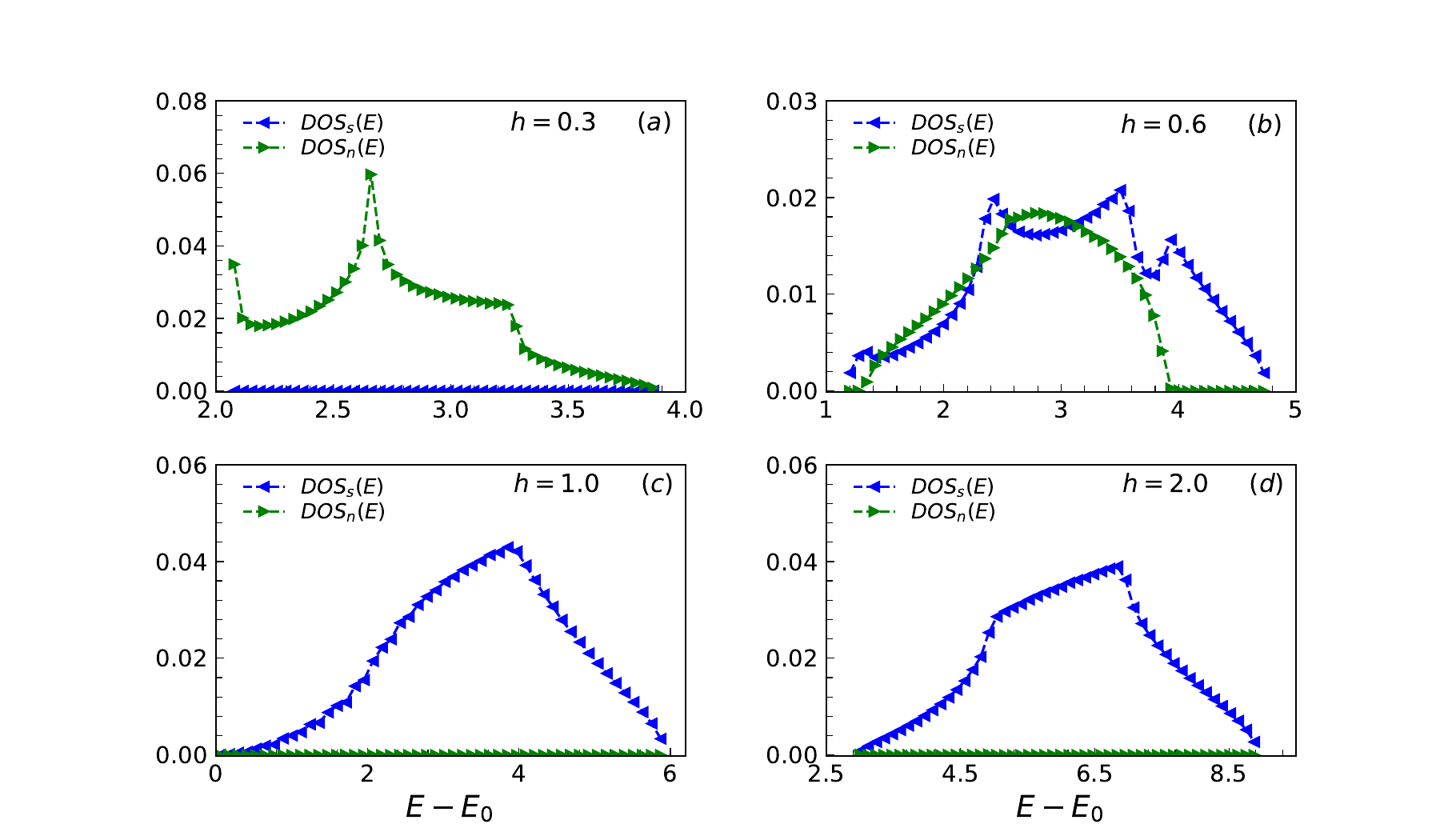}}
\caption{(color online).  The density of squeezed and unsqueezed states as a function of excited energies in the subspace with $N_B=3$ and a chain with size $N=200$. Different values of TF are considered as (a) $h=0.3$, (b) $h=h_f=0.6$, (c) $h=h_c=1.0$, and (d) $h=2.0$.}
 \label{Fig4}
\end{figure}

In our pursuit to better comprehend the nature of the excited states in the model, we have introduced the notion of the density of squeezed and unsqueezed states in a subspace $N_B=m$.  For each value of $N_B=m$, there are $\frac{N!}{m! (N-m)!}$  possible ways to arrange the fermions in the excited states. The energy bands formed by these states are indexed by $N_B = m$ and have a width of $ \Delta = E_{max} - E_{min}$, with the SSP  computed for each state. 
Subsequently, we partition this width into $m'$ equal parts, denoted as $\Delta'=\frac{\Delta}{m'}$.  Finally, we define the density of the squeezed $DOS_s(E)$, and unsqueezed $DOS_n(E)$, states  as
\begin{eqnarray}\label{eq14}
DOS_s(E)&=&\frac{m! (N-m)!}{N!}~N_{s}(E), \nonumber \\
DOS_n(E)&=&\frac{m! (N-m)!}{N!}~N_{n}(E),
\end{eqnarray}
where $N_{s}(E)$ and $N_{n}(E)$ are, respectively, the number of squeezed and unsqueezed states in the interval $E+\Delta'$. We also look at the distribution of the squeezing.  
To achieve this, we partition the width of the squeezed range into $m\rq{}\rq{}$ equal parts and then count the number of squeezed $N'_{s} (\xi_s^2)$, and unsqueezed $N'_{n} (\xi_s^2)$, states within each part. The distribution of squeezed $D_s(\xi_s^2)$, and unsqueezed $D_n(\xi_s^2)$, states is represented by 
\begin{eqnarray}\label{eq15}
D_s(\xi_s^2)&=&\frac{m!(N-m)!}{N!}~N'_{s} (\xi_s^2), \nonumber \\
D_n(\xi_s^2)&=&\frac{m!(N-m)!}{N!}~N'_{n} (\xi_s^2).
\end{eqnarray}

The results on the density of squeezed (unsqueezed) states are graphically illustrated in Fig.~\ref{Fig4} for a chain size $N=200$ and different values of  TF ($m'=50$ is chosen).   We focus on the subspace with $N_B=3$ and Fig.~\ref{Fig4} shows the results. For TF below the factorized field, there are no squeezed states in the spectrum, as shown in Fig.~\ref{Fig4}(a). The low excited states are more unsqueezed than the high excited states and the density of unsqueezed states peaks at the middle of the spectrum. At the factorized field, $h_f=0.6$, about half of the excited states in the subspace become squeezed, as shown in Fig.~\ref{Fig4}(b). The density of squeezed states is higher at the middle of the spectrum than at the edges. The density of unsqueezed states has a similar behavior, except that the high excited states are only squeezed. Fig.~\ref{Fig4}(c) and Fig.~\ref{Fig4}(d) show the results at the quantum critical point $h_c=1$ and a field above the quantum critical point, $h=2.0$, respectively. In these cases, there are no unsqueezed states in the spectrum. The density of squeezed states reaches a maximum at the middle of the spectrum.
\begin{figure}
\centerline{\includegraphics[width=1.15\linewidth]{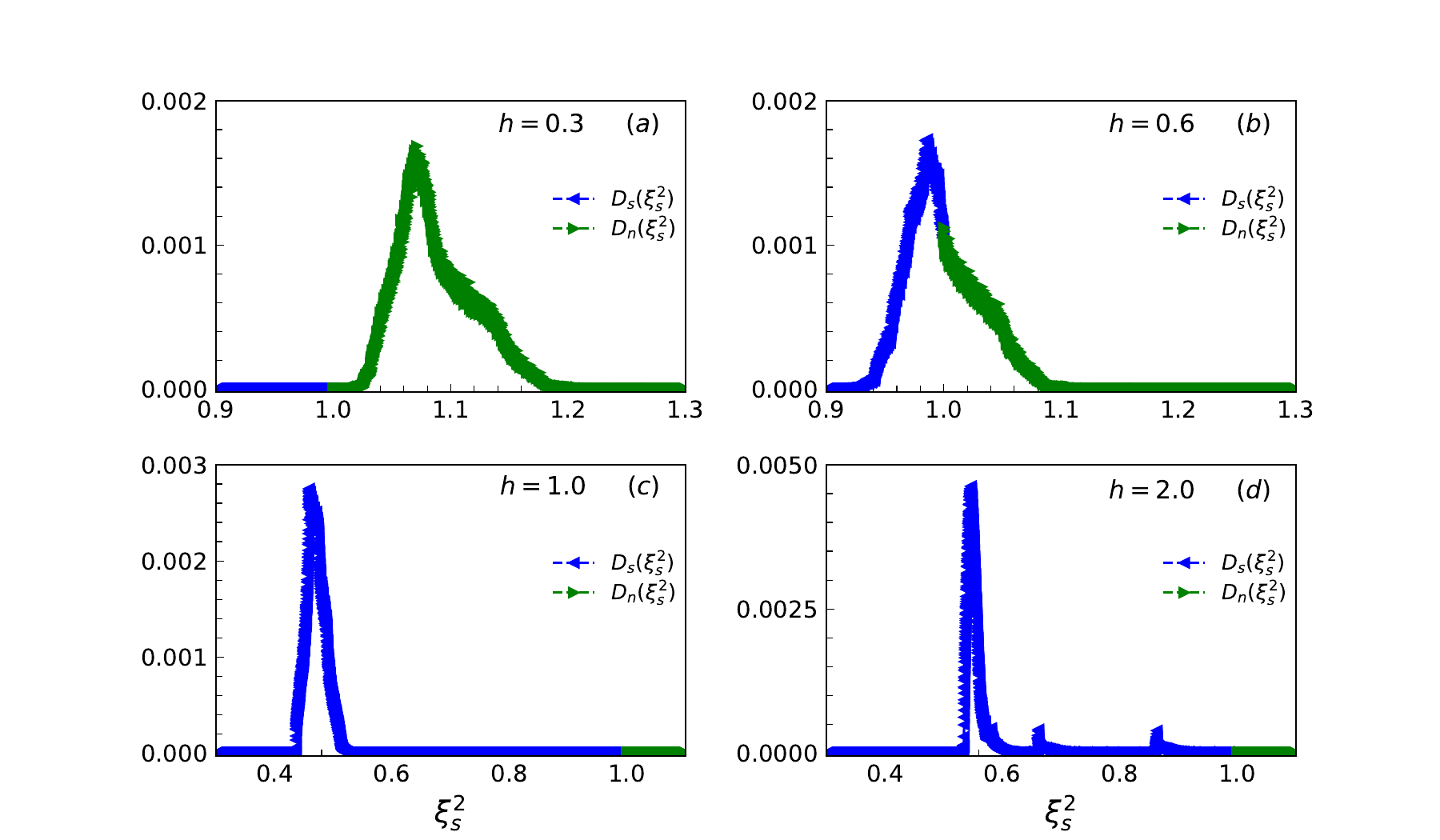}}
\caption{(color online). The distribution of squeezing on its spectrum in the subspace with $N_B=3$ and a chain with size $N=200$. Different values of TF are considered as (a) $h=0.3$, (b) $h=h_f=0.6$, (c) $h=h_c=1.0$, and (d) $h=2.0$. }
\label{Fig5}
\end{figure}
Our results can be understood by using the idea of typicality, which says that most of the states in a big Hilbert space have similar features, such as energy, entropy, and entanglement [{\color{blue}\onlinecite{E20-1}}]. Typical states are the ones that make up the majority of the states in the Hilbert space. For a quantum spin system, typical states have an energy close to the average energy, which is in the middle of the energy spectrum. We can estimate the density of squeezed and unsqueezed states in the energy spectrum by looking at the SSP of typical states. Our results show that typical states are squeezed or unsqueezed in general. This means that the density of squeezed and unsqueezed states is also high in the middle of the energy spectrum. This is because typical states are complex and random, and have a lot of information in their correlations. This result applies to generic quantum systems that do not have any special symmetries or restrictions that would make them less complex.

To complete our study, we analyze the distribution of squeezing ($m\rq{}\rq{}=5000$ is chosen) in a chain of size $N=200$ and the same fields as in Fig.~\ref{Fig4}. We show the results in Fig.~\ref{Fig5}. The most squeezed and unsqueezed states are located in the middle of the SSP nonzero range. Figure.~\ref{Fig5}(a) displays that for TF below the factorized field, the SSP nonzero range is in the unsqueezed region ($\xi^{2}_{s}>1$). As TF increases, the SSP nonzero range shifts to the squeezed region and overlaps with it at the factorized field, $h_f=0.6$. For higher TF, such as the quantum critical point, $h_c=1$, the SSP nonzero range is completely in the squeezed region. For TF above the quantum critical point, the squeezed region becomes narrower and new squeezed regions appear (Fig.~\ref{Fig5}(d)).

\begin{figure}
\centerline{\includegraphics[width=1.1\linewidth, height=0.22\textheight]{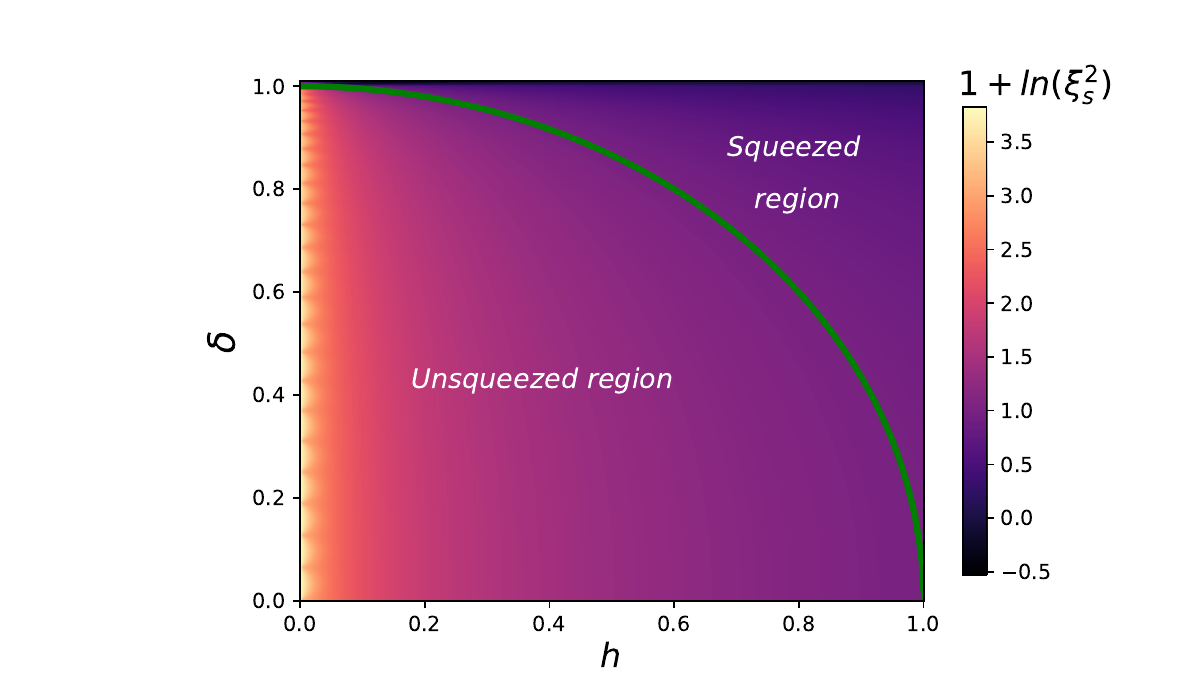}}
\caption{(color online). The density plot of the SSP versus $h$ and $\delta$ at zero temperature for a chain size $N=200$. The green line is $h_f^2+\delta ^2=1$  where the coherent state is located, with $\xi_s^2=1.0$.}
\label{Fig6}
\end{figure}

\section{conclusion}

We have explored the temperature impact on the SSP in an exactly solvable model, i.e., a 1D TF XY model with spin-1/2.  The eigenstates of the Hamiltonian and the total Bogoliubov number operator are the same because they commute. The vacuum state of the Bogoliubov number operator is the ground state of the system and the excited states are in subspaces with a fixed eigenvalue of the Bogoliubov number operator.

This model has a well-known ground state with two phases at zero temperature. For $h<h_c=1$, there is a FM phase with order and for $h>h_c$, there is a PM phase with the disorder. There is also a special point, $h_f^2+\delta^2 = 1$, where the ground state is an eigenstate of the total spin's $z$-component. This point is called the factorized point in which the ground state is a coherent state. Previous studies have shown that the factorized point can act as a boundary point between two regions in this model  [{\color{blue}\onlinecite{E24,E19-1,Dutta, Mahd-1}}]. 
In addition, in the phase diagram of the model, the mentioned line marks a crossover in the FM phase, known as the disordered line, that separates the commensurate and incommensurate phases. The SSP is also an aspect that is connected to the factorization point.  The ground state phase diagram is divided into two regions at zero temperature: $h <h_f$ with no spin squeezing, and $h >h_f$ with spin squeezing.  To show it explicitly, in Fig.~\ref{Fig6} we have displayed the $\xi _s^2$ against $h$ and $\delta$ at zero temperature for a system with size $N=200$. Here the green line indicates the factorized line, $\xi _s^2=1.0$.  At $\delta=0$, the model transitions into the spin-1/2 $XX$ model, which exhibits a Luttinger-liquid phase at zero temperature. In this phase, it has been demonstrated that the SSP can attain very large values [{\color{blue}\onlinecite{20n-2, 20nn}}]. 

We here have investigated the thermal behavior of the SSP by introducing the thermal factorized field and coherent temperature. Our results unveiled that for $h<h_f(T_{co})$, there is no temperature transition. However, at the thermal factorized point, a transition from an unsqueezed thermal state to a squeezed thermal state occurs. Further, we indicated that in such a case where the system sits at the crossover line at zero temperature, the finite temperature can build squeezed thermal states from coherent states.

To gain a better understanding of the thermal behavior of the SSP, we have defined two statistical functions the density of the squeezed and unsqueezed excited states and the distribution of $\xi _s^2$ in the excited states. We have performed a detailed analysis of these statistical functions related to the thermal behavior of the SSP.

The experimental realization of spin squeezing in a one-dimensional transverse field XY model with spin-1/2 can be achieved using ultracold atomic gases trapped in optical lattices [{\color{blue}\onlinecite{Exp-1,Exp-2}}]. These systems provide a highly controllable environment where the parameters of the Hamiltonian, such as the transverse field and interaction strengths, can be precisely tuned. Recent advancements in quantum simulation techniques have demonstrated the feasibility of engineering spin models with high fidelity. For instance, by employing a combination of laser cooling and trapping techniques, it is possible to prepare a one-dimensional chain of spin-1/2 particles. The transverse field can be introduced using a magnetic field or through Raman transitions induced by laser fields. Additionally, the temperature of the system can be controlled by adjusting the cooling parameters, allowing for the exploration of thermal effects on spin squeezing. Experimental setups have successfully observed spin squeezing and related phenomena in similar systems, providing a solid foundation for the realization of the theoretical predictions discussed in this paper.

\acknowledgments
This work was supported by Kuwait University, Research Grant No. SP01/22.

\setcounter{equation}{0}
\renewcommand\theequation{A\arabic{equation}}
\section*{Appendix}
As mentioned, the mean-spin direction is along the $z$ direction which guarantees $\langle {J_x } \rangle  = \langle {J_y } \rangle  = 0$. Consequently, we also obtain $\langle {J_\eta J_z} \rangle  = \langle {J_z J_\eta } \rangle  = 0$ with $\eta=x,y$.
On the other side, $J_{\vec{n}_{\perp}} = \cos (\Omega )J_x + \sin (\Omega)J_y$ that leads to
\begin{eqnarray}\label{eqa1}
({\Delta {J_{\vec{n}_{\perp}}}})^2 &=& \left\langle (J_ {\overrightarrow n  \bot } )^2 \right\rangle  - \left\langle J_ {\overrightarrow n  \bot }  \right\rangle ^2 \nonumber \\
&=& \langle (\cos(\Omega)J_x + \sin(\Omega)J_y)^2\rangle,
\end{eqnarray}
By entering this into Eq.~(\ref{eq6}) one finds that
\begin{eqnarray} \label{eqa2}
\xi _s^2 &=& \frac{2}{N}\mathop {\min }\limits_\Omega  \Big( {\left\langle {J_x^2 + J_y^2} \right\rangle } + \cos (2\Omega )\left\langle {J_x^2 - J_y^2} \right\rangle \nonumber \\ 
& & \ \ \ \ \ \ \ \ \ \ \ \ + \sin(2\Omega)\langle J_xJ_y + J_yJ_x\rangle \Big) \nonumber \\
& = & \frac{2}{N}\Big( {\left\langle {J_x^2 \!+\! J_y^2} \right\rangle  \!-\! \sqrt {{{\left\langle {J_x^2 \!-\! J_y^2} \right\rangle }^2} \!+\! {{\left\langle {{J_x}{J_y} \!+\! {J_y}{J_x}} \right\rangle }^2}} } \Big). \nonumber \\   
\end{eqnarray}
With transitional symmetry of the chain, one can define correlation functions $G_n^{pq} := \langle S_1^{p} S_{1+n}^{q}\rangle$ and thus obtain
\begin{equation} \label{eqa3}
\langle {J_p }{J_q}\rangle = N\langle {S_1^p S_1^q }\rangle  + N\sum\limits_{n = 1}^{N - 1} G_n^{pq }, \ \ \ p,q = x, y, z
\end{equation}
This yields a closed expression for the SSP given by
\begin{multline} \label{eqa4}
\xi _s^2 =  1+ 2\sum\limits_{n = 1}^{N - 1} ( G_n^{xx} + G_n^{yy}) \\ 
- 2\sqrt {\big[\sum\limits_{n = 1}^{N - 1}(G_n^{xx}- G_n^{yy}) \big] ^2 +\big[\sum\limits_{n = 1}^{N - 1}( G_n^{xy} + G_n^{yx}) \big]^2 }.  
\end{multline}
It is simple to write $G_n^{pq}$ in the form of  strings of operators shown as
\begin{eqnarray}\label{eqa5}
G_n^{xx} &=& \frac{1}{4}\langle {{B_1}{A_2}{B_2} \cdots {A_n}{B_n}{A_{n + 1}}}\rangle, \\
G_n^{yy} &=& \frac{{{{( - 1)}^n}}}{4}\langle {{A_1}{B_2}{A_2} \cdots {B_n}{A_n}{B_{n + 1}}} \rangle, \nonumber\\
G_n^{xy} &=& \frac{i}{4}\langle {{B_1}{A_2}{B_2} \cdots {A_n}{B_n}{B_{n + 1}}}\rangle, \nonumber\\
G_n^{yx} &=& \frac{{i{{( - 1)}^{n - 1}}}}{4}\langle {{A_1}{B_2}{A_2} \cdots {B_n}{A_n}{A_{n + 1}}} \rangle, \nonumber
\end{eqnarray}
 where $A_j=c_j^\dag + c_j$, $B_j=c_j^\dag - c_j$. 
In this situation, one can read $G ^{pq}_n$ in the generic form versus Pfaffians $G_n^{pq } = D_n^{pq }\langle {{\phi _1}{\phi _2}{\phi _3} \cdots {\phi _{2n - 2}}{\phi _{2n - 1}}{\phi _{2n}}} \rangle $  where
\begin{eqnarray} \label{eqa6}
 G_n^{pq } = D_n^{pq }\mbox{pf} \left( {\begin{array}{*{20}{c}}
{\left\langle {{\phi _1}{\phi _2}} \right\rangle }&{\left\langle {{\phi _1}{\phi _3}} \right\rangle }&{\left\langle {{\phi _1}{\phi _4}} \right\rangle }& \cdots &{\left\langle {{\phi _1}{\phi _{2n}}} \right\rangle }\\
{}&{\left\langle {{\phi _2}{\phi _3}} \right\rangle }&{\left\langle {{\phi _2}{\phi _4}} \right\rangle }& \cdots &{\left\langle {{\phi _2}{\phi _{2n}}} \right\rangle }\\
{}&{}&{\left\langle {{\phi _3}{\phi _4}} \right\rangle }& \cdots &{\left\langle {{\phi _3}{\phi _{2n}}} \right\rangle }\\
{}&{}&{}& \ddots & \vdots \\
{}&{}&{}&{}&{\left\langle {{\phi _{2n - 1}}{\phi _{2n}}} \right\rangle }
\end{array}} \right). \nonumber
\end{eqnarray}
As we see, there is no situation in which  $n=m$, hence, according to Eq.~(\ref{eq13}), $\langle A_n A_m \rangle =\langle B_n B_m \rangle=0$, and as a result, $G_n^{xy}=G_n^{yx}=0$. This helps us to write a compact form of $\xi _s^2 $ displayed in Eq.~(\ref{eq7}).


\end{document}